\documentclass[
 reprint,
 amsmath,amssymb,
 aps,
]{revtex4-1}

\pdfoutput=1

\usepackage{graphicx}
\usepackage{dcolumn}
\usepackage{bm}

\usepackage{url}
\usepackage{caption}
\begin{document}


\title{Real-time markerless tumour tracking with patient-specific deep learning using a personalized data generation strategy:\\ Proof of concept by phantom study}

\author{Wataru Takahashi}
 \email{tkwataru@shimadzu.co.jp}
\author{Shota Oshikawa}%
\affiliation{%
 Technology Research Laboratory, Shimadzu Corporation, Kyoto, 619-0237, Japan
}%


\author{Shinichiro Mori}
 \email{mori.shinichiro@qst.go.jp}
\affiliation{
 Research Center for Charged Particle Therapy, National Institute of Radiological Sciences, Chiba, 263-8555, Japan
}%


\date{\today}

\begin{abstract}
\begin{description}
\item[Objective]
For real-time markerless tumour tracking in stereotactic lung radiotherapy, we propose a different approach which uses patient-specific deep learning (DL) using a personalized data generation strategy, avoiding the need for collection of a large patient data set. We validated our strategy with digital phantom simulation and epoxy phantom studies.
\item[Methods]
We developed lung tumour tracking for radiotherapy using a convolutional neural network trained for each phantom's lesion by using multiple digitally reconstructed radiographs (DRRs) generated from each phantom's treatment planning 4D-CT. We trained tumour-bone differentiation using large numbers of training DRRs generated with various projection geometries to simulate tumour motion. We solved the problem of using DRRs for training and X-ray images for tracking by using the training DRRs with random contrast transformation and random noise addition.
\item[Results]
We defined adequate tracking accuracy as the \% frames satisfying \textless 1 $\mathrm{mm}$ tracking error of the isocentre. In the simulation study, we achieved 100\% tracking accuracy in 3-$\mathrm{cm}$ spherical and $1.5 \times 2.25 \times 3$-$\mathrm{cm}$ ovoid masses. In the phantom study, we achieved 100\% and 94.7\% tracking accuracy in 3- and 2-$\mathrm{cm}$ spherical masses, respectively. This required 32.5 $\mathrm{ms/frame}$ (30.8 $\mathrm{fps}$) real-time processing.
\item[Conclusions]
We proved the potential feasibility of a real-time markerless tumour tracking framework for stereotactic lung radiotherapy based on patient-specific DL with personalized data generation with digital phantom and epoxy phantom studies.
\item[Advances in Knowledge]
Using DL with personalized data generation is an efficient strategy for real-time lung tumour tracking.
\end{description}
\end{abstract}

\maketitle


\section{\label{sec:level1}INTRODUCTION}

Image-guided radiotherapy (IGRT) is commonly used in commercial stereotactic radiotherapy systems for focusing irradiation on a tumour subject to motion. The technology to locate lung tumour position in real time during irradiation is important because of the trend towards real-time adaptive treatment. Some commercial stereotactic radiotherapy systems support real-time tracking of gold fiducial markers\cite{1SyncTraX}\cite{2Harada}\cite{3Shiinoki}. However, marker implantation is invasive and carries the risks of pneumothorax and marker migration\cite{4Bhagat}. Markerless tracking is being investigated for the next generation of IGRT\cite{5Bahig}\cite{6Yang}\cite{7Dhont}. In most cases, template matching\cite{8Patel}\cite{9Shiinoki}\cite{10Teske} or a correlation model\cite{11Shieh} with X-ray images as a training data set are used. Other cases use a correlation model\cite{12Li} with digitally reconstructed radiographs (DRRs) as training data sets. These methodologies involve the generation of a small personalized training data set for each patient. However, conventional template matching and correlation models often cause robustness problems due to inter- and intra-fractional change or induced artefacts in computed tomography.

Image recognition by machine learning, ng trackespecially deep learning (DL), is another strategy to improve the robustness of markerless tracking\cite{13Terunuma}. DL is a de facto standard for robust image recognition methods. However, DL for medical imaging usually requires multiple-subject data sets for training. Further, data collection is challenging and does not always work well due to the heterogeneity of patient data.

\begin{figure}[htb]
  \begin{center} 
    \includegraphics[width=8cm]{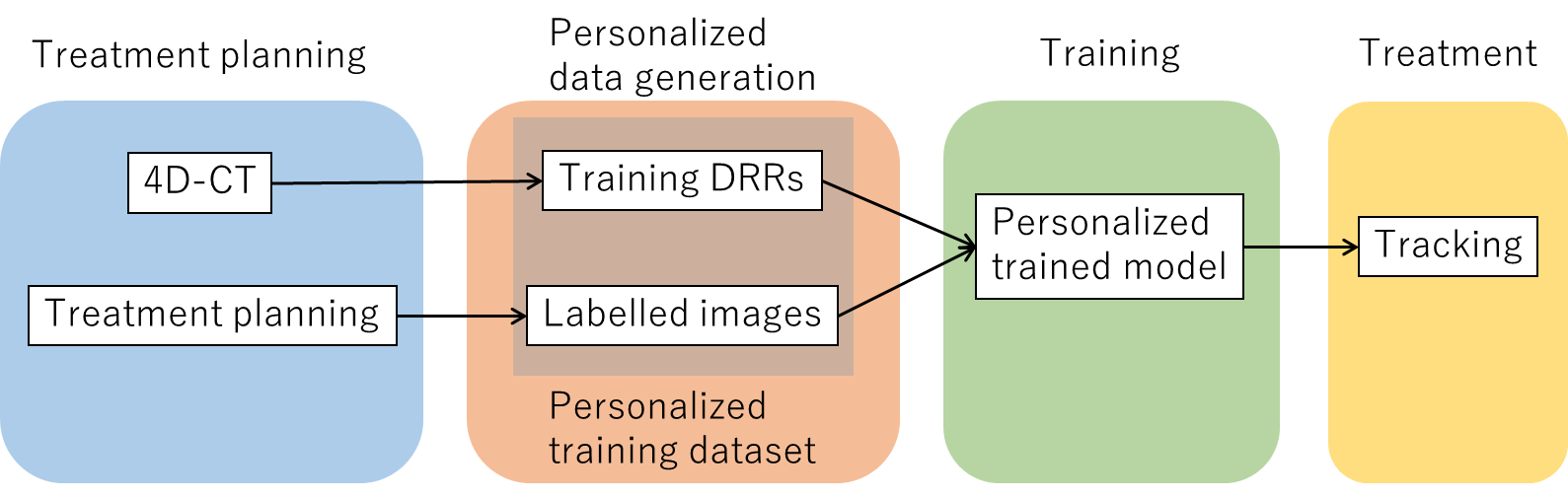}
    \caption{\label{fig:epsart}Procedure for markerless tumour tracking by patient-specific DL in treatment workflow.} 
    \label{fig1} 
  \end{center}
\end{figure}

Here, we propose a different strategy using patient-specific DL with large personalized training data sets generated from individual patients for real-time markerless lung tumour tracking, avoiding the need for collection of a large patient data set. Our strategy uses a personalized training data set generated from each patient's 4-dimensional treatment planning computed tomography (4D-CT). \textbf{FIG. \ref{fig1}} shows tumour tracking by patient-specific DL in treatment workflow. The personalized data generation and training process could be done end-to-end automatically using treatment planning with 4D-CT data. Moreover, these processes could be done in the treatment facility without taking patient data out of the facility, avoiding privacy problems. The tracking is performed during treatment, with the possibility of pretreatment rehearsal.

We validated the feasibility of our strategy by evaluatiing accuracy in both a digital phantom simulation study and an epoxy phantom study.

\section{\label{sec:level1}MATERIALS AND METHODS}

\subsection{\label{sec:level2}Deep learning for markerless tracking}

\begin{figure}[htb]
  \begin{center} 
    \includegraphics[width=8cm]{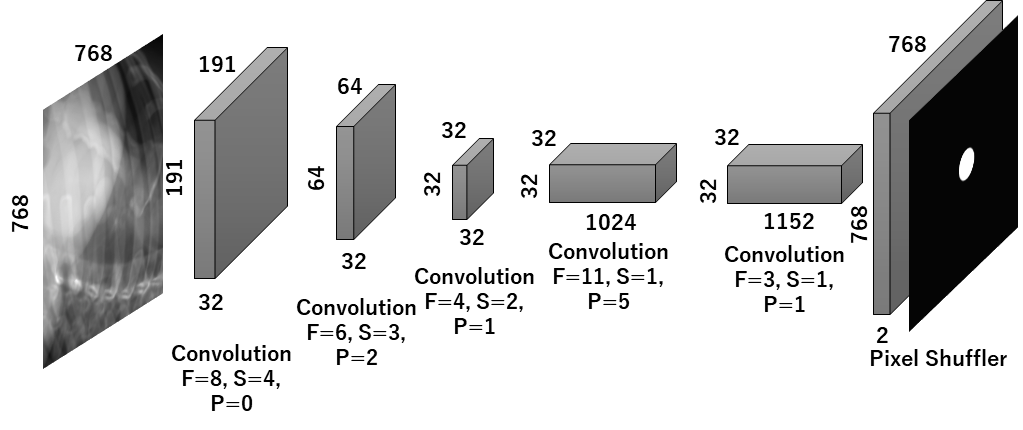}
    \caption{\label{fig:epsart}The neural network model based on FCN combined with a pixel shuffle layer. F: Filter size. S: Stride width. P: Padding size. The left-side image is an input image and the right-side image is a labelled image classified to tumour area (white) or background area (black).} 
    \label{fig2} 
  \end{center}
\end{figure}

We designed a neural network model (\textbf{FIG. \ref{fig2}}) for 2-class semantic segmentation based on a fully convolutional neural network (FCN)\cite{14FCN}. Our model classifies an input image to tumour area or background area pixel-wise for markerless tumour tracking. We combined the FCN with a pixel shuffle layer\cite{15PixShuf} instead of deconvolution layers for \textgreater 15 $\mathrm{fps}$ (\textless 66.7 $\mathrm{ms/frame}$) real-time processing. We used wider convolution sizes in our model than typical FCN using deconvolution layers such as the original FCN and U-net\cite{16U-Net}, because we needed not only local textures but also non-local features such as tumour contours to recognize the tumour. A pixel shuffle layer is faster than deconvolution layers, making it suitable for real-time processing.

In the training process, we used DRRs generated from each phantom's 4D-CT as input images. Also, we used labelled images classified to tumour area or background area as teacher images. We explained the detail of training data set, namely, pairs of a training DRR and a labelled image, in the simulation study section and the phantom study section (below). We trained models by using softmax cross entropy as a loss function and using Adam (adaptive moment estimation)\cite{17Adam} as an optimization algorithm in 200 mini batches and 10 epochs. We consider that Adam optimization is suitable for stable and fast training for a variety of data sets without tuning training parameters for each data set, because Adam automatically updates a learning rate parameter in its internal algorithm. Each projection view was trained independently by using each training data set.

\subsection{\label{sec:level2}The simulation study with a digital 4D-CT phantom}
\subsubsection{Digital 4D-CT phantom}

In the simulation study, a digital respiratory motion phantom including ribs and vertebrae in the form of a 4D-CT (XCAT\textregistered, Duke University, Durham NC, USA)\cite{18XCAT}\cite{19XCAT_HP} was used. The XCAT consists of $512 \times 512 \times 400$ voxels (1 $\mathrm{mm^3}$ voxel) and features 200 phases between peak inhalation and peak exhalation.

We created a digital tumour motion phantom synchronized with the XCAT respiratory motion with the same spatial resolution. The digital tumour phantom motion range was 40 $\mathrm{mm}$ in the superior-inferior (SI) direction, the main component of respiratory motion. We created 3-$\mathrm{cm}$ spherical and $1.5 \times 2.25 \times 3$-$\mathrm{cm}$ ovoid digital tumour phantoms with CT values of $100$ Hounsfield units (HU), which are large enough to be overlapped by ribs.

\subsubsection{Training data sets}

We used only ten phases with the same phase interval with the XCAT digital phantom, because 4D-CT generally consists of ten phases. DRRs were generated by projecting each 4D-CT phase with the same projection geometry as the two-view fluoroscopic tracking systems in the epoxy phantom study. DRRs were $768 \times 768$ pixels ($388 \times 388$ $\mu m^2$ pixels), identical to the flat panel detector (FPD) in the epoxy phantom study (below). The DRR generation algorithm was based on a ray tracing algorithm\cite{20Siddon} and was also used for patient positioning with image registration\cite{21Mori_DRR}\cite{22Mori_first}.

\begin{figure}[htb]
  \begin{center} 
    \includegraphics[width=8cm]{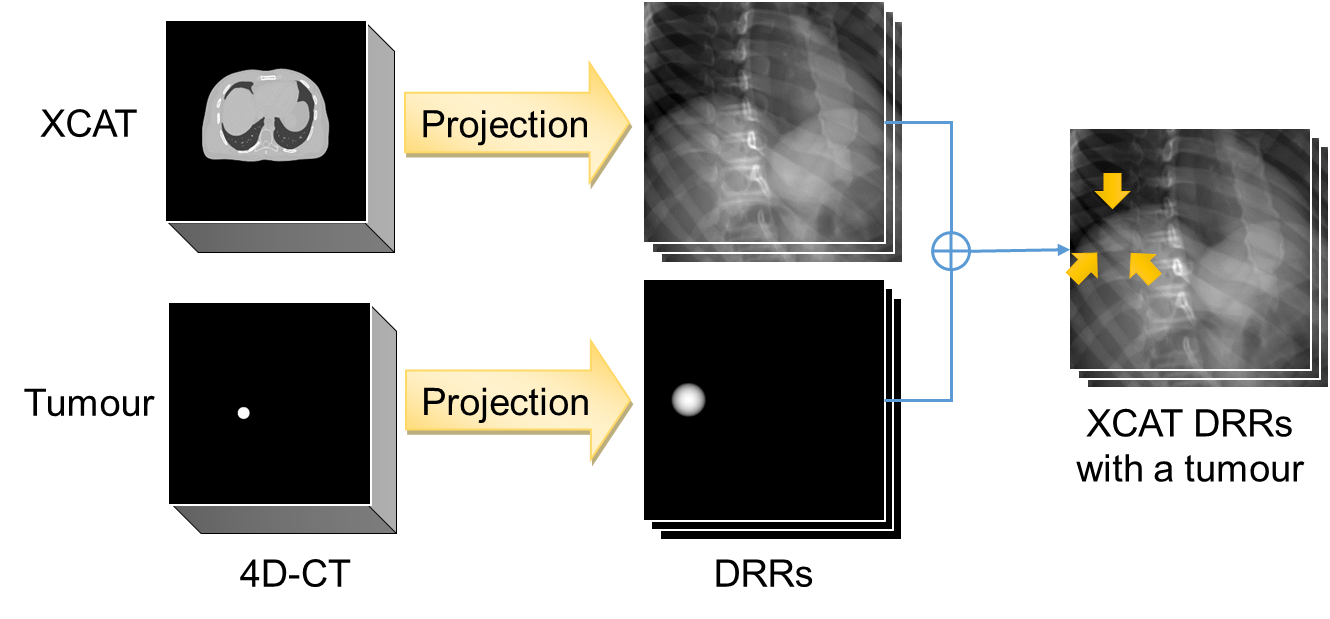}
    \caption{\label{fig:epsart}Generating XCAT digitally reconstructed radiographs (DRRs) with a tumour as training DRRs from XCAT and a digital motion tumour phantom on 4D-CT.} 
    \label{fig3} 
  \end{center}
\end{figure}

We coupled an XCAT DRR and a tumour DRR using 4D-CT with the same projection geometry, allowing us to acquire an XCAT DRR with a tumour as a training DRR (\textbf{FIG. \ref{fig3}}). 

We solved the problem of overlapping bone by using large numbers of training DRRs generated with displacing projection geometries to simulate a range of tumour motion. These training DRRs used the digital tumour phantom with overlapping bone. This solution also works as data augmentation.

\begin{figure*}
\begin{ruledtabular}
  \begin{tabular}{ccc} 
    \rotatebox[origin=c]{90}{\textbf{View 1}} &
    \begin{minipage}{0.49\linewidth}
      \centering
      \scalebox{0.25}{\includegraphics{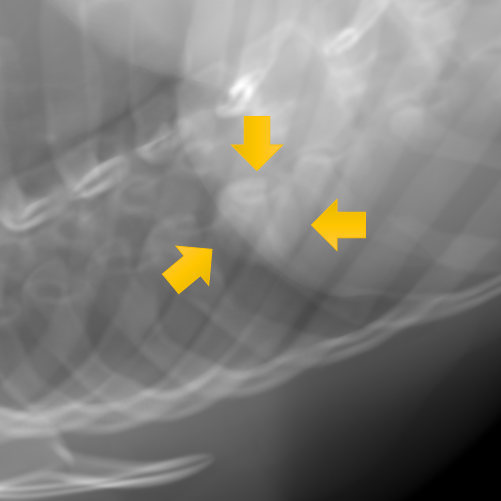}}
    \end{minipage} &
    \begin{minipage}{0.49\linewidth}
      \centering
      \scalebox{0.25}{\includegraphics{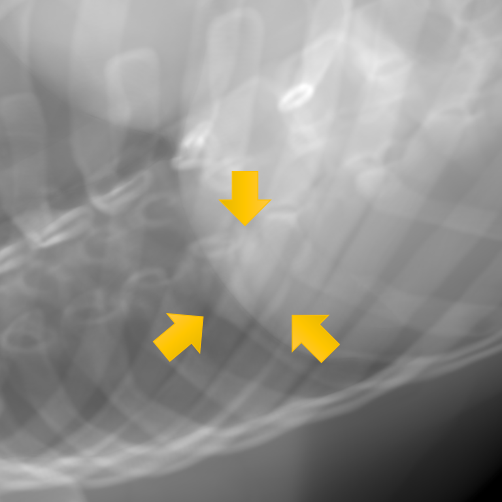}}
    \end{minipage} \\
    & $(x, y, z) = (+6, +6, +6),$ & $(x, y, z) = (-6, -6, -6),$ \\
    & $(\psi, \varphi, \theta) = (+2, +2, +2)$ & $(\psi, \varphi, \theta) = (-2, -2, -2)$\\ \hline \hline
    \rotatebox[origin=c]{90}{\textbf{View 2}} &
    \begin{minipage}{0.49\linewidth}
      \centering
      \scalebox{0.25}{\includegraphics{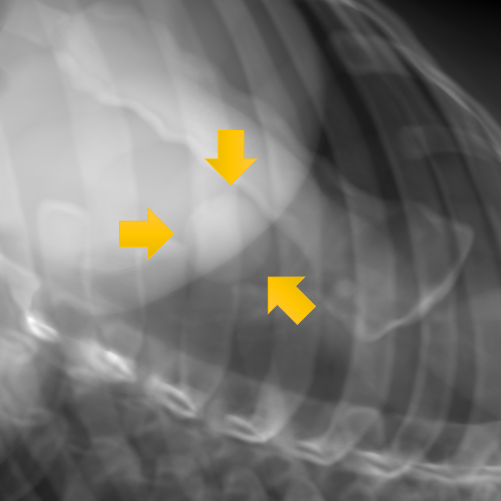}}
    \end{minipage} &
    \begin{minipage}{0.49\linewidth}
      \centering
      \scalebox{0.25}{\includegraphics{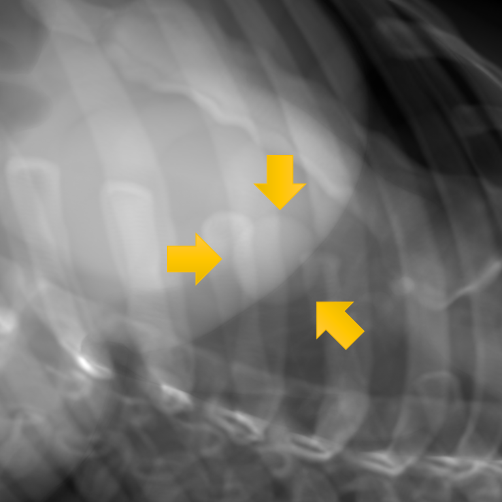}}
    \end{minipage} \\
    & $(x, y, z) = (+6, +6, +6),$ & $(x, y, z) = (-6, -6, -6),$ \\
    & $(\psi, \varphi, \theta) = (+2, +2, +2)$ & $(\psi, \varphi, \theta) = (-2, -2, -2)$\\
  \end{tabular}
  
  \caption{\label{fig:epsart}Examples of training digitally reconstructed radiographs (DRRs) of a spherical tumour phantom with displaced projection geometry using digital phantoms. ($x$, $y$, $z$) and ($\psi$, $\varphi$, $\theta$) are displacements from the original projection geometry.}
  \label{fig4} 

  \vspace{4mm}

  \begin{tabular}{ccc} 
    \rotatebox[origin=c]{90}{\textbf{View 1}} &
    \begin{minipage}{0.49\linewidth}
      \centering
      \scalebox{0.25}{\includegraphics{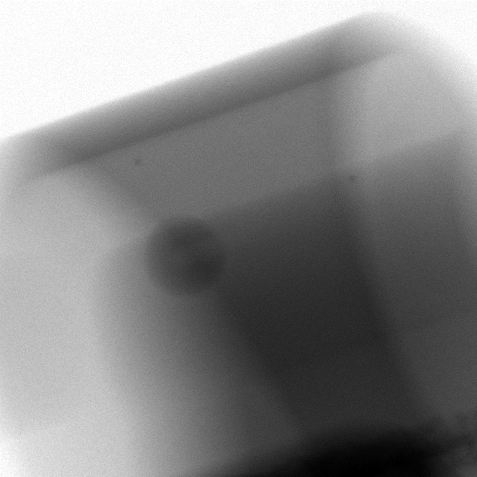}}
    \end{minipage} &
    \begin{minipage}{0.49\linewidth}
      \centering
      \scalebox{0.25}{\includegraphics{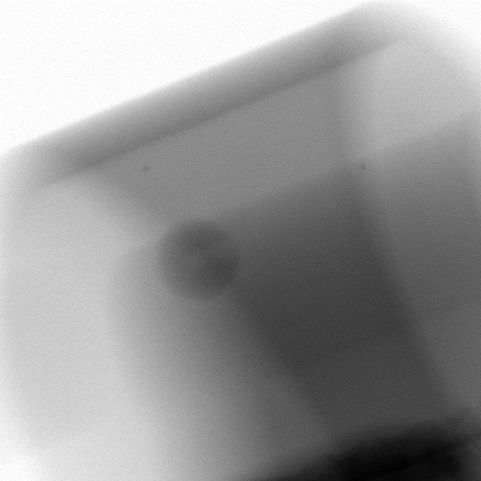}}
    \end{minipage} \\
    & $(x, y, z) = (+6, +6, +6),$ & $(x, y, z) = (-6, -6, -6),$ \\
    & $(\psi, \varphi, \theta) = (+2, +2, +2)$ & $(\psi, \varphi, \theta) = (-2, -2, -2)$\\ \hline \hline
    \rotatebox[origin=c]{90}{\textbf{View 2}} &
    \begin{minipage}{0.49\linewidth}
      \centering
      \scalebox{0.25}{\includegraphics{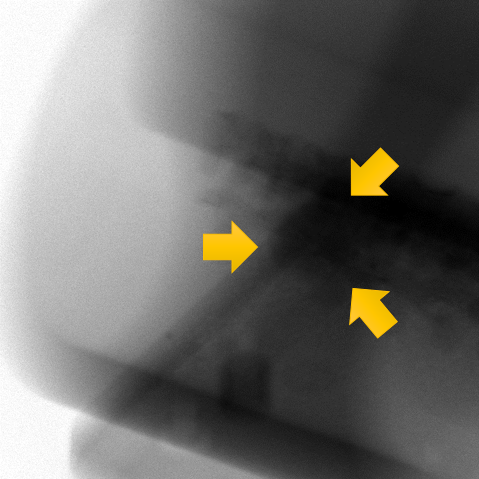}}
    \end{minipage} &
    \begin{minipage}{0.49\linewidth}
      \centering
      \scalebox{0.25}{\includegraphics{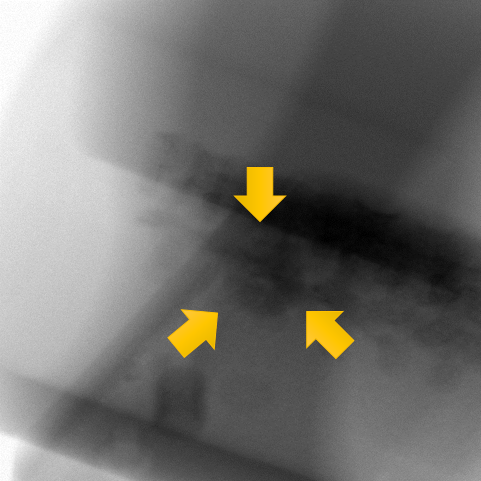}}
    \end{minipage} \\
    & $(x, y, z) = (+6, +6, +6),$ & $(x, y, z) = (-6, -6, -6),$ \\
    & $(\psi, \varphi, \theta) = (+2, +2, +2)$ & $(\psi, \varphi, \theta) = (-2, -2, -2)$\\
  \end{tabular}
  \caption{\label{fig:epsart}Examples of training digitally reconstructed radiographs (DRRs) of a 3-cm tumour phantom with displaced projection geometry, random contrast transformation, and random noise. ($x$, $y$, $z$) and ($\psi$, $\varphi$, $\theta$) are displacements from the base projection geometry.}
  \label{fig5} 

\end{ruledtabular}
\end{figure*}

To generate the DRRs, we displaced projection geometry with 3-dimensional translation ($x$, $y$, $z$) and rotation ($\psi$, $\varphi$, $\theta$) of an X-ray tube/FPD, by $-6$ to $+6$ $\mathrm{mm}$ with 3-$\mathrm{mm}$ steps and $-2^\circ$ to $+2^\circ$ with $1^\circ$ steps. We then acquired $5^6 - 1 = 15,624$ DRRs per 4D-CT phase (\textbf{FIG. \ref{fig4}}). Finally, we acquired $156,240$ DRRs in ten 4D-CT phases. Because we can interpolate and extrapolate tumour motion in discrete 4D-CT phases, the geometric displacements were calculated from the tumour phantom motion range between 4D-CT phases.

We also generated large numbers of labelled images corresponding to training DRRs. In a labelled image, tumour area was calculated as the projection area of a tumour in 4D-CT. The other area was labelled as background area. This allowed us to automatically generate phantom-specific training data sets, namely pairs of a training DRR and a labelled image.

\subsubsection{Markerless tracking}

DRRs for all 4D-CT phases with the same projection geometry, without displacements, were used as tracking images. These tracking images consisted of 200 frames and were not included in the training data set. 

A tumour coordinate was calculated from semantic segmentation of a tracking image by using a phantom-specific trained model for every frame. Trained models output the `tumour score' (0.0-1.0) pixel-wise. Therefore, we were able to calculate a tumour coordinate as the density centre 
weighted by pixel-wise tumour score thresholding \textgreater 0.5. Using the density centre potentially allows acquisition of a tumour coordinate with sub-pixel accuracy in a tracking image. Misdetections of a few pixels had little effect on the density centre coordinate. Therefore, we expected higher accuracy than that of object detection methods outputting an object coordinate directly, such as Faster R-CNN\cite{23Ren}.

\subsubsection{Tracking accuracy evaluation}

The ground truth of a tumour coordinate is the density centre of a tumour DRR weighted by pixel values with sub-pixel accuracy. We evaluated `tracking error' as the 2-dimensional distance between a tracking coordinate and a ground truth coordinate in each view. We evaluated not only absolute error on the isocentre in mm but also relative error on FPD by pixel units, because tracking errors may be affected by pixel resolution. We defined `tracking accuracy' as the ratio of frames satisfying \textless 1 $\mathrm{mm}$ tracking error on the isocentre. Tracking accuracy corresponds to gating accuracy in typical isotropic PTV-to-CTV margins of 1 $\mathrm{mm}$\cite{24Mori_track}. In case with no pixels having a \textgreater 0.5 tumour score, we considered that tumour detection did not succeed in this frame.

\subsection{\label{sec:level2}The phantom study with an epoxy respiratory motion phantom}
\subsubsection{Epoxy respiratory motion phantom}

In the phantom study, we validated multi-modality DL, which uses DRRs for training and X-ray images for tracking, using an epoxy respiratory motion phantom. A programmable respiratory motion phantom (Dynamic Thorax Phantom MODEL 008A\textregistered, Computerized Imaging Reference Systems, Inc., Norfolk VA, USA) \cite{25CIRS} was used. This is an epoxy chest phantom with vertebrae but no ribs. It allows the operator to change spherical tumour size (1 $\mathrm{cm}$, 2 $\mathrm{cm}$ and 3 $\mathrm{cm}$ diameter). The CT values of the epoxy `tumours' are approximately 30 $\mathrm{HU}$.

A 4D-CT of the MODEL 008A phantom was acquired (Aquilion ONE\textregistered, Canon Medical Systems Corporation, Otawara City, Japan) with different sized tumour phantoms. We marked phantom position with the CT scanner's laser markers to reposition the chest phantom for tracking in fluoroscopy. We then guided a tumour phantom on a rod in a sinusoidal motion ($\pm20$ $\mathrm{mm}$ amplitude and 4 second cycle) in the SI direction. 4D-CT phases were synchronized with tumour motion. This 4D-CT consisted of $512 \times 512 \times 204$ voxels ($0.625 \times 0.625 \times 1$ $\mathrm{mm}$ voxel) and ten phases for a full single respiratory cycle (T00T90). Because 4D-CT was acquired in the same manner as during treatment workflow, in discrete phases, the 4D-CT imaging range of tumour motion was approximately 38 $\mathrm{mm}$, which was narrower than the actual tumour motion range. Also, motion artefacts, which gave the tumours a more strongly differentiated appearance than most cases of inter- and intra-fractional changes, were detected in some 4D-CT phases.

\subsubsection{Training data sets}

We generated large numbers of training DRRs with displacing projection geometry from 4D-CT data of the chest phantom to simulate a variety of tumour motion patterns. We displaced the projection geometry by $-6$ to $+6$ $\mathrm{mm}$ with 3-$\mathrm{mm}$ steps and $-2^\circ$ to $+2^\circ$ with $1^\circ$ steps, respectively. We then acquired $5^6 = 15,625$ DRRs per 4D-CT phase (\textbf{FIG. \ref{fig5}}). Finally, we acquired $156,250$ DRRs in all ten 4D-CT phases. Because we can interpolate and extrapolate tumour motion in discrete 4D-CT phases, the geometry displacement range was decided by the tumour motion range between 4D-CT phases. 

In multi-modality DL using DRRs for training and X-ray images for tracking, we faced the specific problem of differences in image quality between DRRs and X-ray images, caused mainly by beam hardening effect, scattering, sensitivity characteristics of FPDs, and shot noise. Therefore, a DRR cannot be converted to an X-ray image precisely, because pixel values of a DRR do not correspond to pixel values of an X-ray image one-to-one. We solved this problem by using large numbers of training DRRs with random contrast transformation and random noise addition.

The contrast variation range was decided by analysing differences in contrast between DRRs and X-ray images. We applied gamma transformation with random $\gamma$ values ($\gamma = 0.4–2.5$) to each DRR as random contrast transformation. We also added Gaussian noise ($\sigma = 25$) as random noise such that the noise variation range was $\pm1$ standard deviation of the X-ray image noise.

Just as in the simulation study, we also generated large numbers of labelled images corresponding to training DRRs.

\subsubsection{Markerless tracking}

\begin{figure}[htb]
  \begin{center} 
    \includegraphics[width=8cm]{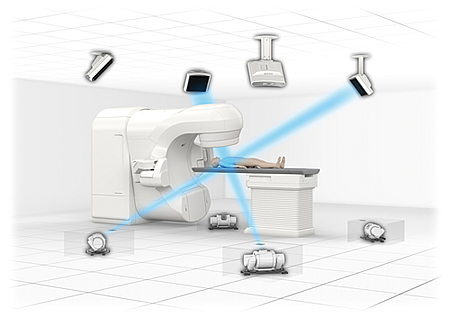}
    \caption{\label{fig:epsart}Geometry of the two-view X-ray fluoroscopy imaging system SyncTraX FX4\textregistered with a radiotherapy system. Two views cross at the isocentre without being blocked by the gantry head. Source-object distance (SOD) = 2353 $\mathrm{mm}$. Source-image distance (SID) = 4172 $\mathrm{mm}$.} 
    \label{fig6} 
  \end{center}
\end{figure}

Two-view X-ray fluoroscopy images of the chest phantom were simultaneously taken and marked with the laser positioning system (SyncTraX FX4\textregistered, Shimadzu Corporation, Inc., Kyoto, Japan) \cite{1SyncTraX} (\textbf{FIG. \ref{fig6}}). We manually positioned the phantom as close as possible to its location in the 4D-CT using the laser markers. Because we had not yet constructed a synchronizing system between X-ray fluoroscopy frame and measurement of tumour position, we acquired X-ray images at the 25 fixed tumour positions on the same trajectory taken by the 4D-CT with a tumour phantom inserted. The other structures of the phantom except the `tumour' were fixed at the same position. The X-ray settings were 110 $\mathrm{kV}$, 50 $\mathrm{mA}$, 4 $\mathrm{ms}$, the same as in our gold marker tracking. These X-ray images consisted of $768 \times 768$ pixels ($388 \times 388$ $\mu m^2$ pixel). They were taken in slightly different positions than in the 4D-CT because of differences in phase intervals.

We applied a Gaussian blur filter to each X-ray image to simulate the same spatial resolution as training DRRs, which was lower than that of X-ray images, because DRR spatial resolution was limited by the CT spatial resolution (image quality of DRRs and X-ray images was close to each other for multi-modality DL). These blurred X-ray images were used as tracking images.

As in the simulation study, a tumour coordinate was calculated from semantic segmentation of a tracking image by using a phantom-specific trained model for every frame. A tumour coordinate was calculated as the density centre of the tumour score with sub-pixel accuracy, which was essentially unaffected by the Gaussian blur filter.

\subsubsection{Tracking accuracy evaluation}

\begin{figure}[htb]
  \begin{center} 
    \includegraphics[width=8cm]{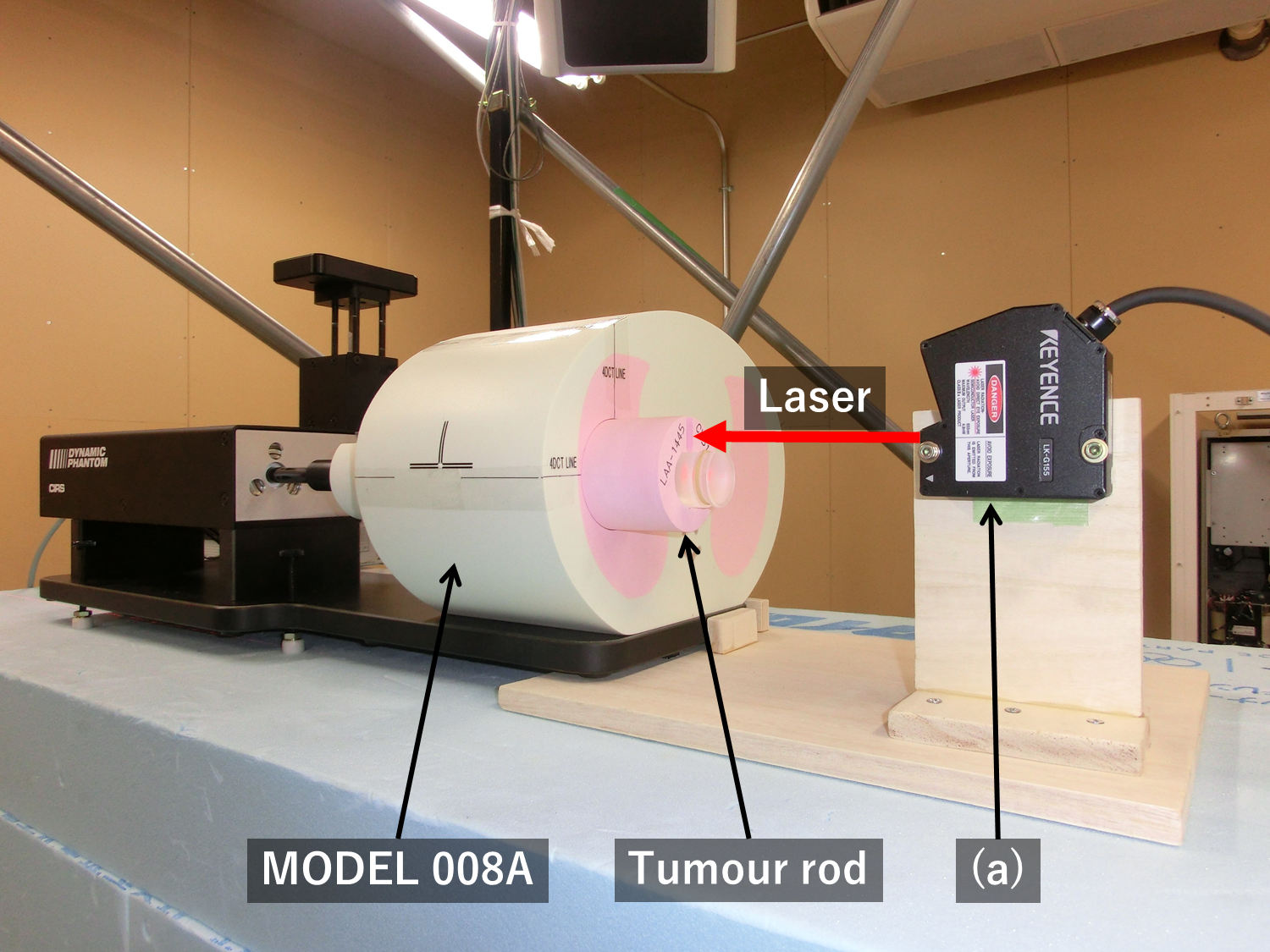}
    \caption{\label{fig:epsart}Tracking accuracy evaluation system using a MODEL 008A phantom with a laser displacement meter (a). We measured displacement of a motion tumour rod of the MODEL 008A phantom outside both X-ray fields of view.} 
    \label{fig7} 
  \end{center}
\end{figure}

We measured displacement of a motion tumour rod of MODEL 008A phantom as ground truth of the tumour coordinate using a laser displacement meter (LK-G155\textregistered, Keyence Inc., Osaka, Japan, \textless 20 $\mu m$ accuracy) outside of the X-ray fields of view (\textbf{FIG. \ref{fig7}}). We had previously calibrated between displacements of a tumour phantom and tumour coordinates on X-ray images. We then acquired tumour coordinates from displacements of the tumour phantom at 25 different positions. 

As in the simulation study, we evaluated not only absolute error on the isocentre in mm but also relative error on FPD by pixel units. Also, we evaluated `tracking accuracy'.

\subsection{\label{sec:level2}Implementation}

The personalized data generation process was programmed with C++ and CUDA\textregistered (NVIDIA Corporation, Santa Clara CA, USA) running on the GPU. The training process and the tracking process were programmed with Python with the open-source DL framework Chainer (Preferred Networks, Inc., Tokyo, Japan), running on the GPU.

All processes (i.e., personalized data generation, training, and tracking) were done with a single computer (Windows\textregistered 7 64 bit, Intel Xeon CPU E5-2637 dual, 64GB RAM, Microsoft, Inc. Redmond WA, USA; NVIDIA Quadro M6000 24GB GPU dual). Personalized data generation, training, and tracking were run as two-view processes simultaneously by using dual GPUs.

\section{\label{sec:level1}RESULTS}
\subsection{\label{sec:level2}Digital simulation study}

\textbf{FIG. \ref{fig8}} shows an example of a tracking image. There were no false detections, extra-detections, outside the tumour area, in spite of many similar rib structures in the tracking image.

\textbf{FIG. \ref{fig9}} shows the tracking error distribution for all frames. The tracking error distribution was \textless 1 $\mathrm{mm}$ on the isocentre and almost \textless 1 pixel on the FPD without effects of pixel resolution in both views and with both tumour shapes.

\textbf{TABLE \ref{tab1}} shows a summary of tracking accuracy. We achieved 100\% tracking accuracy in both views and with both tumour shapes in spite of overlying bone.

\begin{ruledtabular}
\begin{figure*}
  \begin{tabular}{ccc} 
     & \textbf{Spherical tumour} & \textbf{Ovoid tumour}\\ \hline
    \rotatebox[origin=c]{90}{\textbf{View 1}} &
    \begin{minipage}{0.49\linewidth}
      \centering
      \scalebox{0.2}{\includegraphics{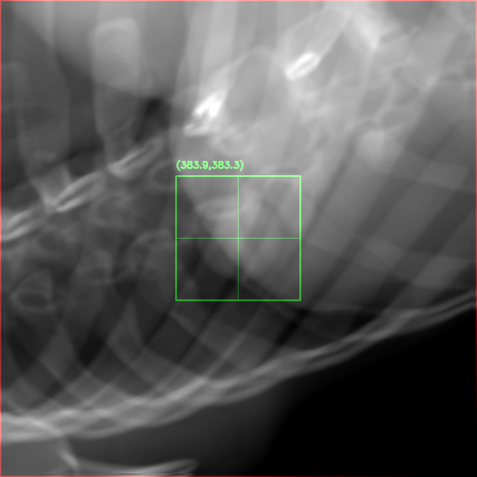}}
    \end{minipage} &
    \begin{minipage}{0.49\linewidth}
      \centering
      \scalebox{0.2}{\includegraphics{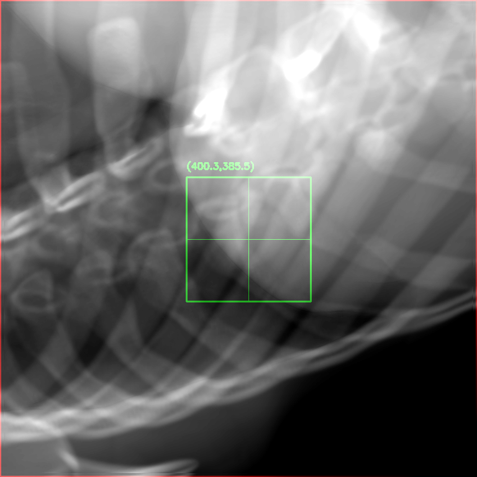}}
    \end{minipage} \\ \hline \hline
    \rotatebox[origin=c]{90}{\textbf{View 2}} &
    \begin{minipage}{0.49\linewidth}
      \centering
      \scalebox{0.2}{\includegraphics{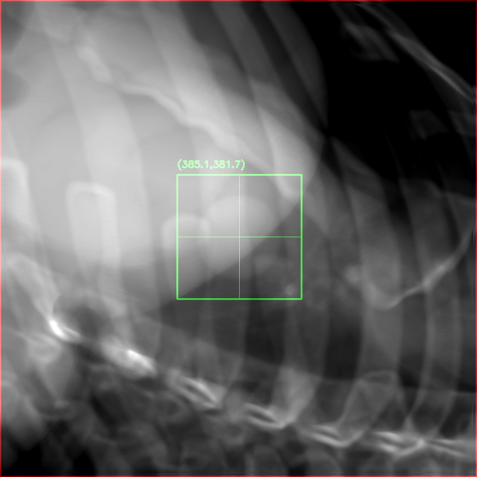}}
    \end{minipage} &
    \begin{minipage}{0.49\linewidth}
      \centering
      \scalebox{0.2}{\includegraphics{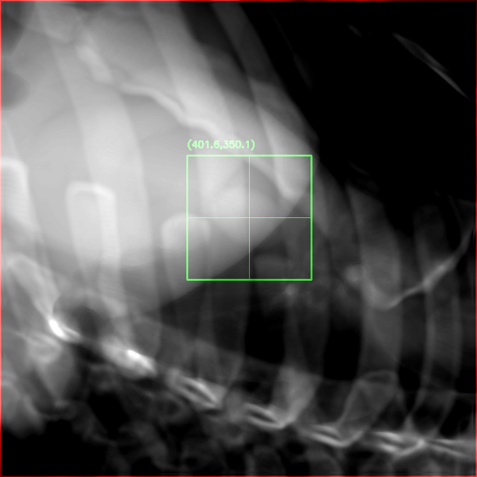}}
    \end{minipage} \\
  \end{tabular}
  \caption{\label{fig:epsart}An example of a tracking image in the digital simulation study. A calculated tumour coordinate is the centre of the green rectangle in the tracking image.}
  \label{fig8} 
\end{figure*}
  
\begin{figure*}
  \begin{tabular}{ccc} 
     & \textbf{Spherical tumour} & \textbf{Ovoid tumou}r\\ \hline
    \rotatebox[origin=c]{90}{\textbf{View 1}} &
    \begin{minipage}{0.49\linewidth}
      \centering
      \scalebox{0.2}{\includegraphics{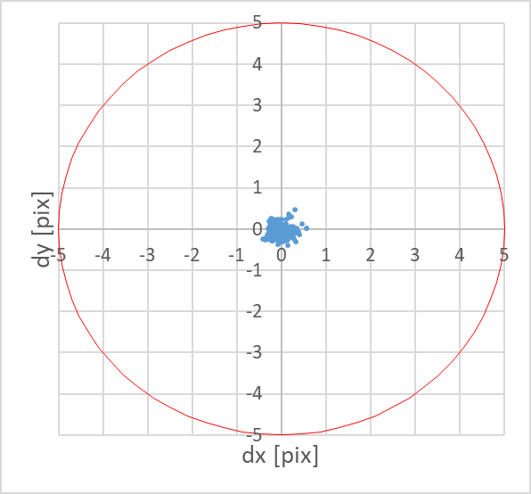}}
    \end{minipage} &
    \begin{minipage}{0.49\linewidth}
      \centering
      \scalebox{0.2}{\includegraphics{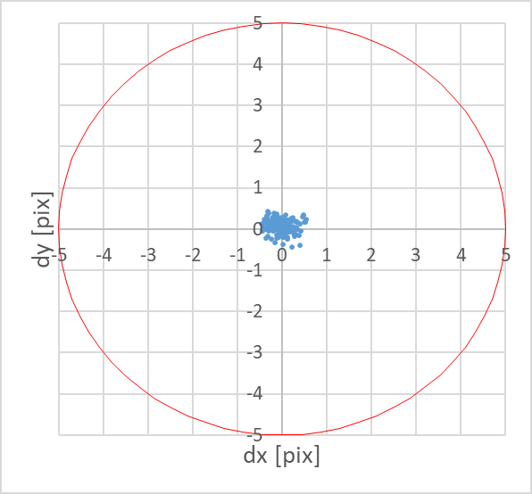}}
    \end{minipage} \\ \hline \hline
    \rotatebox[origin=c]{90}{\textbf{View 2}} &
    \begin{minipage}{0.49\linewidth}
      \centering
      \scalebox{0.2}{\includegraphics{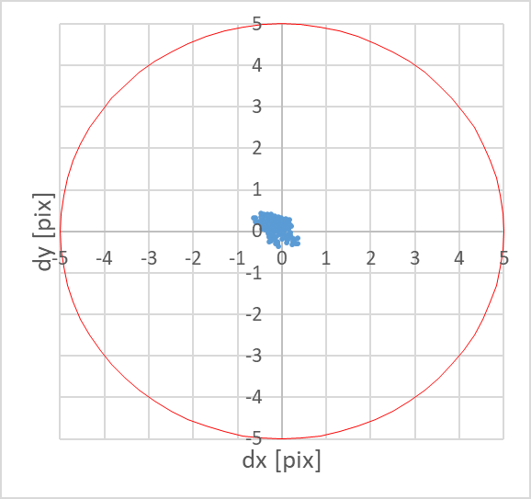}}
    \end{minipage} &
    \begin{minipage}{0.49\linewidth}
      \centering
      \scalebox{0.2}{\includegraphics{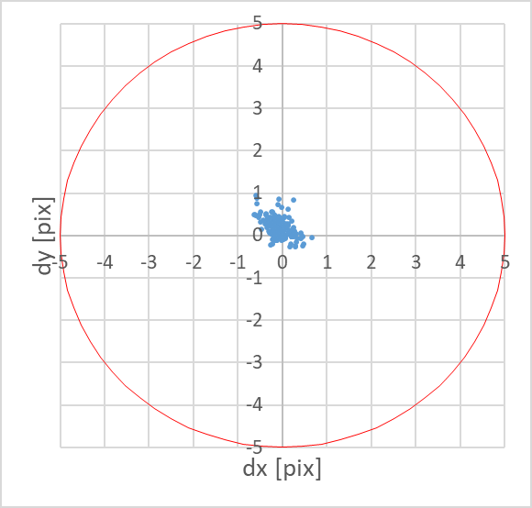}}
    \end{minipage} \\
  \end{tabular}
  \caption{\label{fig:epsart}The 2-dimensional tracking error distribution on FPD for all frames in the digital simulation study. The centre of a graph, the origin, means zero error. The scale units are one pixel on the FPD. The red circle means 1 mm error on the isocentre, the criteria for tracking accuracy.}
  \label{fig9} 
\end{figure*}

\begin{table*}
  \caption{\label{tab:table3}Summary of tracking accuracy in the digital simulation study. Tracking error shows `mean error' $\pm$ `standard deviation' with (maximum error) for all frames.}
\begin{tabular}{cccccc}
  & &\multicolumn{2}{c}{\textbf{View 1}} & \multicolumn{2}{c}{\textbf{View2}}\\
 \multicolumn{2}{c}{\textbf{Tumour shape}} & \textbf{Spherical} & \textbf{Ovoid} & \textbf{Spherical} & \textbf{Ovoid}\\ \hline
 \multicolumn{2}{c}{\textbf{Tracking accuracy}} & 100\% & 100\% & 100\% & 100\%\\ \hline
  & \textbf{Isocentre} & 0.05 $\pm$ 0.02 & 0.05 $\pm$ 0.03 & 0.06 $\pm$ 0.03 & 0.07 $\pm$ 0.04\\
 \textbf{Tracking error} & [$\mathrm{mm}$] & (0.12) & (0.13) & (0.16) & (0.25)\\ \cline{2-6}
 \textbf{(Maximum error)} & \textbf{FPD} & 0.22 $\pm$ 0.11 & 0.23 $\pm$ 0.13 & 0.27 $\pm$ 0.15 & 0.32 $\pm$ 0.20\\
  & [$\mathrm{pixel}$] & (0.56) & (0.60) & (0.72) & (1.13)\\
\end{tabular}
\label{tab1}
\end{table*}
\end{ruledtabular}

\subsection{\label{sec:level2}Epoxy phantom study}

\textbf{FIG. \ref{fig10}} shows an example of a tracking image. There were no false detections in the search area using the 2- and 3-$\mathrm{cm}$ tumours despite the tumour overlapping the spine in view 2. However, there were false detections in the search area in some frames using the 1-$\mathrm{cm}$ tumours.

\textbf{FIG. \ref{fig11}} shows the tracking error distribution for all frames. In the cases of the 2- and 3-$\mathrm{cm}$ tumours, the tracking error distribution was almost \textless 1 mm on the isocentre and almost \textless 5 pixels on the FPD without effects of pixel resolution in both views. However, with the 1-$\mathrm{cm}$ tumours, tracking error distribution was \textgreater 1 mm, and was worse than with the 2- and 3-$\mathrm{cm}$ tumours. A common bias error for all tumour sizes by manual positioning of the phantom was not detected.

\textbf{TABLE \ref{tab2}} shows a summary of tracking accuracy. We achieved \textgreater 94.7\% tracking accuracy in 2- and 3-$\mathrm{cm}$ tumours in spite of multi-modality DL using DRRs for training and X-ray images for tracking. However, tracking accuracy with 1-$\mathrm{cm}$ tumours was only 40.3\%.

\begin{ruledtabular}
\begin{figure*}
  \begin{tabular}{cccc} 
     & \textbf{3-cm tumour} & \textbf{2-cm tumour} & \textbf{1-cm tumour}\\ \hline
    \rotatebox[origin=c]{90}{\textbf{View 1}} &
    \begin{minipage}{0.33\linewidth}
      \centering
      \scalebox{0.2}{\includegraphics{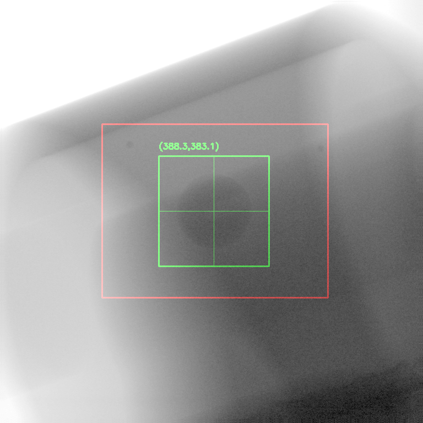}}
    \end{minipage} &
    \begin{minipage}{0.33\linewidth}
      \centering
      \scalebox{0.2}{\includegraphics{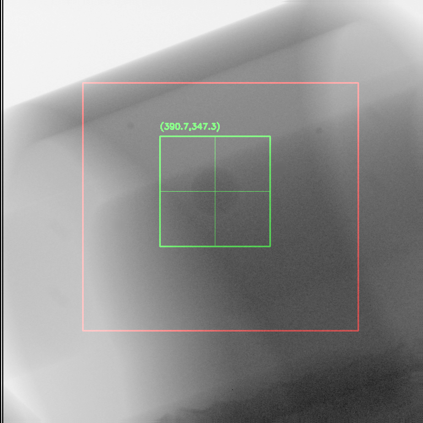}}
    \end{minipage} &
    \begin{minipage}{0.33\linewidth}
      \centering
      \scalebox{0.2}{\includegraphics{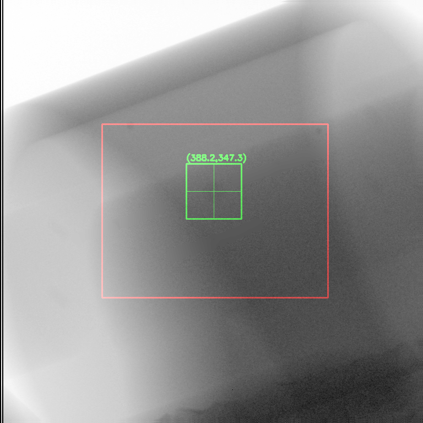}}
    \end{minipage} \\ \hline \hline
    \rotatebox[origin=c]{90}{\textbf{View 2}} &
    \begin{minipage}{0.33\linewidth}
      \centering
      \scalebox{0.2}{\includegraphics{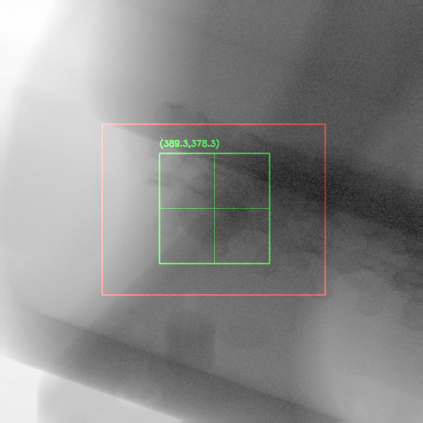}}
    \end{minipage} &
    \begin{minipage}{0.33\linewidth}
      \centering
      \scalebox{0.2}{\includegraphics{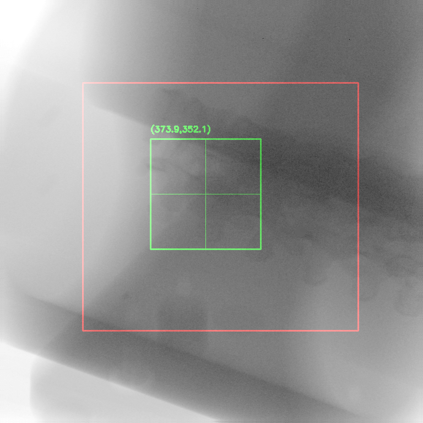}}
    \end{minipage} &
    \begin{minipage}{0.33\linewidth}
      \centering
      \scalebox{0.2}{\includegraphics{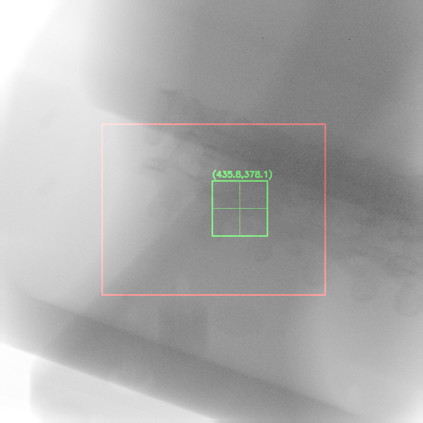}}
    \end{minipage} \\
  \end{tabular}
  
  \caption{\label{fig:epsart}An example of a tracking image in an epoxy phantom study. The calculated tumour coordinate is in the centre of the green rectangle. The red rectangle is the search area for a tumour.}
  \label{fig10} 
\end{figure*}

\begin{figure*}
  \begin{tabular}{cccc} 
     & \textbf{3-cm tumour} & \textbf{2-cm tumour} & \textbf{1-cm tumour}\\ \hline
    \rotatebox[origin=c]{90}{\textbf{View 1}} &
    \begin{minipage}{0.33\linewidth}
      \centering
      \scalebox{0.2}{\includegraphics{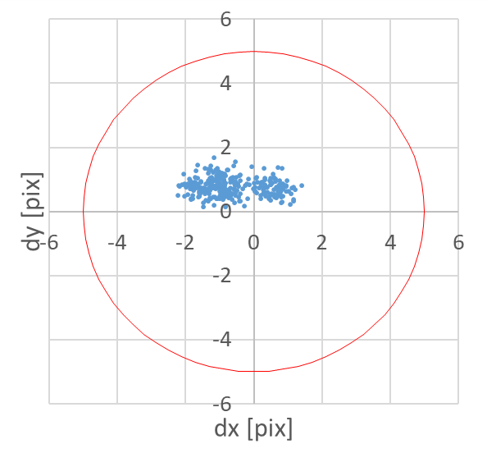}}
    \end{minipage} &
    \begin{minipage}{0.33\linewidth}
      \centering
      \scalebox{0.2}{\includegraphics{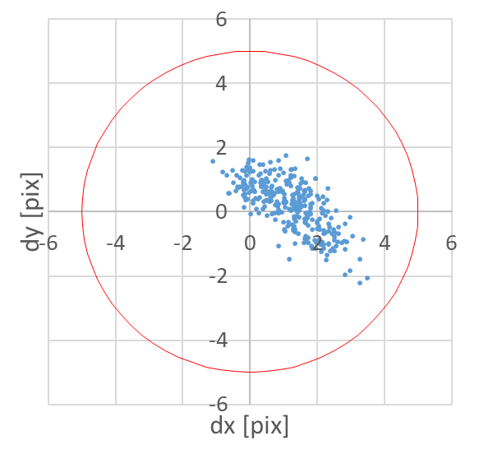}}
    \end{minipage} &
    \begin{minipage}{0.33\linewidth}
      \centering
      \scalebox{0.2}{\includegraphics{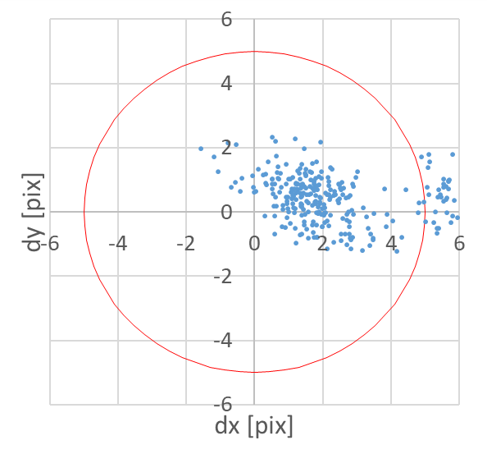}}
    \end{minipage} \\ \hline \hline
    \rotatebox[origin=c]{90}{\textbf{View 2}} &
    \begin{minipage}{0.33\linewidth}
      \centering
      \scalebox{0.2}{\includegraphics{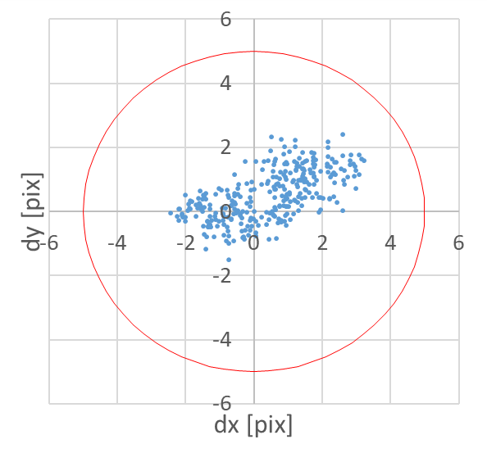}}
    \end{minipage} &
    \begin{minipage}{0.33\linewidth}
      \centering
      \scalebox{0.2}{\includegraphics{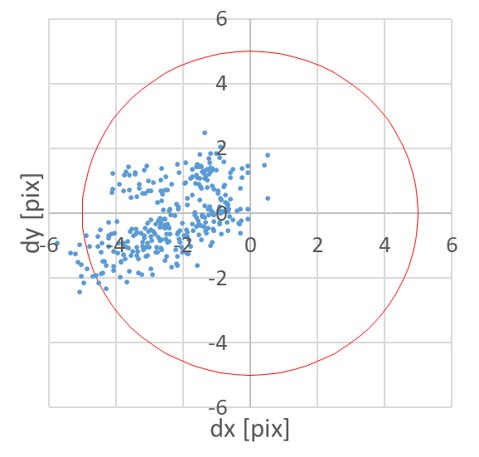}}
    \end{minipage} &
    \begin{minipage}{0.33\linewidth}
      \centering
      \scalebox{0.2}{\includegraphics{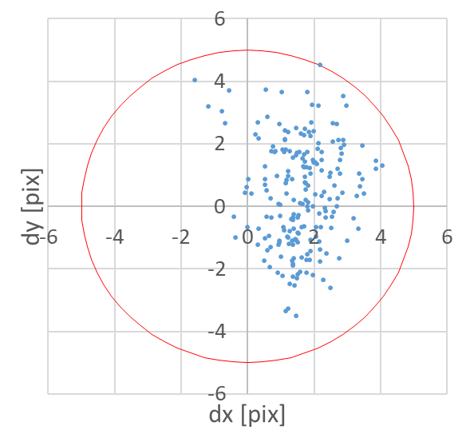}}
    \end{minipage} \\
  \end{tabular}
  
  \caption{\label{fig:epsart}The 2-dimensional tracking error distribution on flat panel detectors (FPD) for all frames (except \textgreater 6-pixel errors) in the epoxy phantom study. The centre of a graph, the origin, means zero error. The scale units are one pixel on FPD. The red circle means 1 $\mathrm{mm}$ error on the isocentre, the criteria of tracking accuracy.}
  \label{fig11} 
\end{figure*}

\begin{table*}
\caption{\label{tab:table3}Summary of tracking accuracy in the epoxy phantom study. Tracking error shows `mean error' $\pm$ `standard deviation' with (maximum error) for all frames.}
\begin{tabular}{cccccccc}
  & &\multicolumn{3}{c}{\textbf{View 1}} & \multicolumn{3}{c}{\textbf{View2}}\\
 \multicolumn{2}{c}{\textbf{Tumour size}} & \textbf{3 cm} & \textbf{2 cm} & \textbf{1 cm} & \textbf{3 cm} & \textbf{2 cm} & \textbf{1 cm}\\ \hline
 \multicolumn{2}{c}{\textbf{Tracking accuracy}} & 100\% & 100\% & 73.1\% & 100\% & 94.7\% & 40.3\%\\ \hline
  & \textbf{Isocentre} & 0.3 $\pm$ 0.1 & 0.3 $\pm$ 0.1 & 0.6 $\pm$ 0.4 & 0.3 $\pm$ 0.2 & 0.5 $\pm$ 0.2 & 1.8 $\pm$ 4.7\\
 \textbf{Tracking error} & [$\mathrm{mm}$] & (0.5) & (0.9) & (1.8) & (0.8) & (1.3) & (30.4)\\ \cline{2-8}
 \textbf{(Maximum error)} & \textbf{FPD} & 1.3 $\pm$ 0.4 & 1.5 $\pm$ 0.6 & 2.9 $\pm$ 1.9 & 1.5 $\pm$ 0.8 & 2.4 $\pm$ 1.1 & 8.4 $\pm$ 21\\
  & [$\mathrm{pixel}$] & (2.4) & (4.0) & (8.3) & (3.6) & (5.8) & (139.8)\\
\end{tabular}
 \label{tab2}
\end{table*}
\end{ruledtabular}

\subsection{\label{sec:level2}Processing time}

The personalized data generation processing took about 9 hours to output $156,250$ DRRs into a hard disk drive (HDD). This processing time is rate-limited by HDD access. The training process took about 17 hours, and is also rate-limited by HDD access. The tracking process (except image reading) took $32.5 \pm 4.7$ $\mathrm{ms/frame}$ (30.8 $\mathrm{fps}$) for real-time processing.

\section{\label{sec:level1}DISCUSSION}

We proved the potential feasibility of a real-time markerless tumour tracking framework for stereotactic lung radiotherapy based on patient-specific DL with a personalized data generation strategy using a digital phantom simulation study and an epoxy phantom study. The personalized data generation and training process could be done end-to-end automatically between treatment planning and treatment, without manual procedures such as template creation just before the treatment process while a patient is positioned and fixed. Also, our framework does not require cone beam CT (CBCT) data taken while a patient is positioned. Therefore, our framework has potential to improve treatment throughput and reduce patient burden, compared to some conventional markerless tracking frameworks which require manual procedures just before the treatment process or require CBCT\cite{8Patel}\cite{9Shiinoki}\cite{10Teske}\cite{11Shieh}. Also, our framework using DL with a large personalized training data set has potential to improve the robustness and accuracy of markerless tracking, compared to frameworks using conventional machine learning with a small personalized training data set\cite{12Li}.

The digital simulation study had 100\% tracking accuracy in spite of overlying bone. Also, as the tracking errors were significantly smaller than the displacements for training DRRs, they could be easily distinguished from each other. These results indicate that our method was able to learn tumour shape while ignoring overlying bone by using training data sets with a variety of overlying bone patterns. Also, we succeeded in tracking both spherical and ovoid tumour phantoms. This result indicates that our method has the potential to track all varieties of tumour shape.

In the epoxy phantom study, we achieved \textgreater 94.7\% tracking accuracy in 2- and 3-$\mathrm{cm}$ tumours in spite of multi-modality DL using DRRs for training and X-ray images for tracking. This result indicates that we were able to allow it to learn modality invariance characteristics by using training data sets with a range of contrast and noise. This was achieved despite discrete 4D-CT phases and a narrower 4D-CT imaging range of tumour motion. This result indicates its success at interpolation and extrapolation of tumour motion. Also, this was achieved without being affected by manual positioning error despite the fixed bone structures. This result indicates that our model did not learn only relative positional relationships between bone structures and a tumour. However, we need additional evaluation of tumour trajectories in all three dimensions, not only in the SI direction, in order to validate its robustness for irregular tumour motion compared with training data. We achieved this accuracy despite motion artefact from a tumour in 4D-CT. We consider that our method has the potential to detect tumours regardless of motion in 4D-CT or inter- and intra-fractional changes in tumour shape. However, we need additional evaluation with a retrospective clinical study to validate its robustness. In 1-$\mathrm{cm}$ tumours, tracking was difficult because of the low contrast of the X-ray images due to scattered radiation. We required additional evaluation with 1-$\mathrm{cm}$ `tumours' in the digital simulation study, which ignores scattered radiation. The tracking accuracy could be improved in these small masses by improving the contrast variation algorithm for training DRRs to simulate scattered radiation. Currently, we have calculated a tumour coordinate using only a present frame without temporal information such as tumour coordinates in past frames. We may be able to reduce tumour detection inaccuracies by using temporal information. In this case, we can evaluate tracking accuracy by using irregular respiratory waveforms. 

In the labelled image, tumour area was labelled as the projection area of a tumour on the 4D-CT. But, in actual treatment workflow, tumour area can be labelled as the planning target volume (PTV) or clinical target volume (CTV) generated by treatment planning by projecting PTV or CTV to a training DRR. In this case, tumour contours also can be acquired by positioning PTV or CTV on the tumour centre coordinate during tracking.

We consider that patient-specific DL has the potential to provide better accuracy for each patient than standard DL using multiple-patient data sets. Patient-specific DL can be considered overfitting for a specific patient. Generally speaking, standard DL provides good accuracy for all patients. Conversely, overfitting provides better accuracy for a specific patient than standard DL.

The personalized data generation process and training process took about 26 hours in total. These processes have to be done between treatment planning and treatment in the workflow. The training time of 26 hours might not be problematic because, generally, treatment commences about one week after treatment planning. However, more than 24 hours may not be sufficient if our method is applied for retraining or fine tuning during the course of treatment, or for instance at an MRI linac. We can substantially shorten these processing times by generating DRRs and training data sets in memory without rate-limiting HDD accessibility.

The tracking process requires \textgreater 15 $\mathrm{fps}$ (\textless 66.7 $\mathrm{ms/frame}$) real-time processing, so our tracking process of 32.5 $\mathrm{ms/frame}$ (30.8 $\mathrm{fps}$) is sufficient.

\section{\label{sec:level1}CONCLUSIONS}

We proved the potential feasibility of a real-time markerless tumour tracking framework for stereotactic lung radiotherapy based on patient-specific DL with personalized data generation by evaluating the tracking accuracy in both digital phantom and epoxy phantom studies.

\section*{\label{sec:level1}CONFLICT OF INTEREST STATEMENT}

Wataru Takahashi and Shota Oshikawa are employees of Shimadzu Corporation, Japan.

\section*{\label{sec:level1}FUNDING}

This work was supported by Shimadzu Corporation, Kyoto, Japan and National Institute of Radiological Sciences, Chiba, Japan.

\begin{acknowledgments}
Construction of the tracking accuracy evaluation system was supported by Ryuichi Ito, Seiji Yamanaka, and Yui Torigoe of Shimadzu Corporation. We thank Libby Cone, MD, MA, from DMC Corp. (\url{http://www.dmed.co.jp/}) for editing drafts of this manuscript.
\end{acknowledgments}

\end{document}